\documentclass[10pt, conference]{IEEEtran}
\IEEEoverridecommandlockouts
\usepackage{cite}
\usepackage{amsmath,amssymb,amsfonts}
\usepackage{algorithmic}
\usepackage{graphicx}
\usepackage{textcomp}
\usepackage{xcolor}
\def\BibTeX{{\rm B\kern-.05em{\sc i\kern-.025em b}\kern-.08em
    T\kern-.1667em\lower.7ex\hbox{E}\kern-.125emX}}

\usepackage{xcolor}
\usepackage{amsmath}
\usepackage{multirow}
\usepackage{subfigure}
\usepackage{tikz}
\usepackage{adjustbox}
\usepackage[normalem]{ulem}
\usepackage{graphicx}
\usepackage{enumitem}
\usepackage[utf8]{inputenc}
\usepackage{titlesec}
\usepackage{lipsum}
\usepackage{booktabs}
\usepackage{mathtools}

\DeclareFixedFont{\ttb}{T1}{txtt}{bx}{n}{8} 
\DeclareFixedFont{\ttm}{T1}{txtt}{m}{n}{8}  
\usepackage{color}
\definecolor{deepblue}{rgb}{0,0,0.5}
\definecolor{deepred}{rgb}{0.6,0,0}
\definecolor{deepgreen}{rgb}{0,0.5,0}

\usepackage{listings}
\newcommand\pythonstyle{\lstset{
language=Python,
basicstyle=\ttm,
morekeywords={self},              
keywordstyle=\ttb\color{deepblue},
emph={test_coffee, config_option_show, resoure_patch, test_abort},          
emphstyle=\ttb\color{deepred},    
stringstyle=\color{deepgreen},
frame=tb,                         
showstringspaces=false
}}
\lstnewenvironment{python}[1][]
{
\pythonstyle
\lstset{#1}
}
{}

\begin{document}

\title{CSA-Trans: Code Structure Aware Transformer for AST}

\author{\IEEEauthorblockN{Saeyoon Oh}
\IEEEauthorblockA{\textit{KAIST} \\
saeyoon17@kaist.ac.kr}
\and
\IEEEauthorblockN{Shin Yoo}
\IEEEauthorblockA{\textit{KAIST} \\
shin.yoo@kaist.ac.kr}
}

\maketitle

\begin{abstract}
When applying the Transformer architecture to source code, designing a good self-attention mechanism is critical as it affects how node relationship is extracted from the Abstract Syntax Trees (ASTs) of the source code. We present Code Structure Aware Transformer (CSA-Trans), which uses Code Structure Embedder (CSE) to generate specific PE for each node in AST. CSE generates node Positional Encoding (PE) using disentangled attention. To further extend the self-attention capability, we adopt Stochastic Block Model (SBM) attention. Our evaluation shows that our PE captures the relationships between AST nodes better than other graph-related PE techniques. We also show through quantitative and qualitative analysis that SBM attention is able to generate more node specific attention coefficients. We demonstrate that CSA-Trans outperforms 14 baselines in code summarization tasks for both Python and Java, while being 41.92\% faster and 25.31\% memory efficient in Java dataset compared to AST-Trans and SG-Trans respectively.
\end{abstract}


\section{Introduction}
The Transformer architecture first proposed by the work of Vaswani~et.~al.~\cite{Vaswani:transformer} for machine translation, has been proven to be powerful in various domains including computer vision~\cite{Dosovitskiy:vit}, natural language processing~\cite{Brown:gpt3}, and chemistry~\cite{Nambiar:proteintrans}. One critical component of the Transformer architecture is self-attention~\cite{Vaswani:transformer} that relates tokens at different positions in the input. It is therefore critical to provide positional information of each token so that Transformer can distinguish tokens at different locations. For sequential inputs, sinusoidal functions have been used to provide absolute PE values to input tokens~\cite{Vaswani:transformer}; later, relative PE was introduced to emphasize relative token distances in sequential inputs~\cite{Shaw:relativeatt}. Various domain specific PE schemes have been proposed for other modalities such as images and graphs.~\cite{Xiangxiang:conditionalpe, Vijay:laplacianpe, Dwivedi:learnablerwpe}. 


Code summarization aims to build brief natural language summary of the given code. Empirical studies show that program comprehension is one of the most time consuming activities for developers during program development~\cite{Minelli:developertime1, Xia:developertime2}. However, documentations and code comments are often outdated, mismatched, or even entirely missing~\cite{Hu:badcomment}. Many techniques have been proposed so that code summaries can be automatically generated~\cite{14McBurney, 10Haiduc, Hu:badcomment, Tang:ast-trans, Gao:sg-trans}.

Yet, most existing work consider source code as sequences of tokens similarly to natural language texts: these models continue to use sequential positional information. Attempts to incorporate AST have been made recently~\cite{Tang:ast-trans,guo:codescribe,Hu:badcomment}, but each have its own limitations. DeepCom~\cite{Hu:badcomment} linearizes ASTs by adopting Structure-Based Traversal (SBT), a specific tree traversal order that can preserve the structural information through brackets. Yet, the information is only implicit, making it difficult for the model to learn the relationship between two arbitrary tokens. Furthermore, SBT traversal generates long linearized sequences, requiring models to learn long range dependencies. AST-Trans~\cite{Tang:ast-trans} incorporates relative distances between AST nodes directly into its attention mechanism. However, the self-attention mechanism only considers predetermined types of node relationship, restricting the receptive field of the self-attention mechanism. \textsc{CodeScribe}~\cite{guo:codescribe} learns PE based on node specific tree positions obtained from AST. While this preserves some structural information, details such as node types are disregarded. Further, the learned PE lacks generality: given an AST with unseen structure, \textsc{CodeScribe} cannot provide PE. This motivates us to propose a novel mechanism for the provision of positional information via a learnable PE that is aware of context information extracted from ASTs.

We present CSA-Trans, a Transformer architecture for source code with PE aware of the context of each node. Our Code Structure Embedder (CSE) produces Code Structure Aware PE (CSA-PE) using the disentangled attention mechanism similar to AST-Trans. Yet, rather than using the node embeddings to generate summary, we incorporate them as node PE. The learned PE is subsequently concatenated to AST embeddings to be fed into the Transformer architecture. With the context information in CSA-PE, CSA-Trans can \emph{learn} how to attend via SBM attention~\cite{Cho:sbmattention} rather than using predefined node pairs for attention calculation. Since CSA-PE does not depend on concrete positions of AST nodes, it is free from the generalization and scalability concerns of \textsc{CodeScribe}.

We evaluate CSA-Trans using both Java and Python code summarization tasks, against 15 baselines (four are manually replicated, while the results for the remaining are taken from the literature~\cite{Gao:sg-trans}). Moreover, we compare different PEs for source code, and show that CSA-PE can learn a good encoding. We devise a synthetic experiment to see how well CSA-PE captures the node relationship compared to other PE schemes. We also present quantitative and qualitative analysis of the learned attention masks and scores compared to the scores from vanilla attention approach, analyzing how SBM attention helps. Finally, we present how efficient CSA-Trans is compared to two strong baselines in terms of time and memory. We include our implementation of CSA-Trans and code for our experiments publically available\footnote{https://github.com/saeyoon17/Code-Structure-Aware-Transformer}.

The remainder of the paper is organized as follows. Section~\ref{sec:preliminary} presents the background information for the Transformer architecture as well as Positional Encodings (PEs), disentangled attention, and the Stochastic Block Model attention mechanism. Section~\ref{sec:csatrans} describes our Transformer architecture, CSA-Trans. Section~\ref{sec:experiments} presents our evaluation, followed by related work in Section~\ref{sec:relatedwork}. Section~\ref{sec:limitations} presents the limitations and Section~\ref{sec:conclusion} concludes.

\section{Preliminary}
\label{sec:preliminary}

\subsection{Transformer}
\label{pre:transformer}
Transformer is an encoder-decoder framework in which each encoder and decoder is consisted of multiple layers. Encoder layer first applies self-attention mechanism upon input sequence. Assume we have $n$ token vectors each of dimension size $d$, $x = [x_1; \dots; x_n] \in \mathbf{R}^{n \times d}$. The self-attention score between $i$th and $j$th token is calculated as follows:

\begin{align}
    \alpha_{ij} =  \frac{Q(x_i)K(x_j)^T}{\sqrt{d}}
\end{align}

\noindent where $Q, K \in \mathbf{R}^{d \times d}$ represent projection matrices. Using the scores, attention output of $i$th token is calculated as:
\begin{align}
    o_i =  \sum_j{\sigma(\frac{\alpha_{ij}}{\sqrt{d}})V(x_j)}
\end{align}

where $V \in \mathbf{R}^{d \times d}$. Vector $O = [o_1, \dots, o_n] \in \mathbf{R}^{N \times d}$ is subsequently fed into the feed-forward network and normalized. For multi-head attention with $h$ heads, $h$ parallel processes with separate $Q, K, V$ matrices are used to generate $O_1, \dots, O_h$, which are concatenated, and multiplied by the weight matrix $W^O$ for the final output.

A similar operation is performed in decoder layers, except for the cross attention. Cross attention scores are calculated between encoder outputs and hidden vectors of the decoder which enables the decoder vectors to learn relevant features from the encoder outputs.

\subsection{Positional Encodings}
Transformer requires the positional information to be either encoded in the input or to be externally given to differentiate tokens at different location. Various encoding schemes throughout different domains have been proposed~\cite{Xiangxiang:conditionalpe, Vighnesh:treepos}. Among them, we introduce four PEs that are applicable for ASTs.

\textbf{Sequential Positional Encoding} (Sequential PE), introduced in the original Transformer~\cite{Vaswani:transformer}, uses sinusoidal functions to encode positions of the sequential tokens. For a given position $pos$ and dimension $i$, the Sequential PE is defined as: 
\begin{align}
    PE_{(pos, 2i)} = sin(pos/f(i))\\
    PE_{(pos, 2i+1)} = cos(pos/f(i))
\end{align}
where $f(i) = 10000^{2i/d}$ and $d$ is the encoding dimension. The encoding is added to input vectors before it is fed into encoder. Incorporating the PE enables model to distinguish tokens at different locations and learn relative positions. However, to use Sequential PE on complex structures, linearization technique is essential. Previous work using Sequential PE either use Pre-Order Traversal~\cite{Iyer:CodeNN, Eriguchi:Tree2Seq} or Structure-Based Traversal (SBT)~\cite{Hu:badcomment} for AST linearization.

\textbf{Tree positional encoding}  (Tree PE)~\cite{Vighnesh:treepos} aims to provide structural positional encoding for tree structured data. It assumes a regular tree with degree $d$ and maximum tree depth $h$ where $d, h$ are hyperparameters. Tree PE stacks child indices (i.e., index $i$ for the $i$th node among its siblings) of each node while traversing from given node to root node. A learnable Tree PE stacks weighted child indices while traversing: the weights for different tree levels are learned. In our evaluation, we use learnable version of Tree PE.

\textbf{Node triplet positions} (Triplet PE)~\cite{guo:codescribe} Each node is given a triplet of (1) depth of the node, (2) width position of its parent, and (3) width position of itself. Subsequently, each distinct triplet position is given a separate learnable embeddings, which is then used as the PE for each AST node.

\textbf{Graph Laplacian Eigenvectors} (Laplacian PE) was suggested by~\cite{Vijay:laplacianpe} to be used as PE for nodes in graphs. It has been shown that the Laplacian PE improves performance of GNNs in various tasks~\cite{Wang:laplacianusage1, Dwivedi:learnablerwpe}. Given a graph and its adjacency $A$, graph Laplacian eigenvectors $U$ is calculated as:
\begin{align}
    \Delta = I - D^{-1/2}AD^{-1/2} = U^TAU
\end{align}
Laplacian PE is known to build good local encodings while preserving global structures, serving as meaningful node PEs.

\subsection{Disentangled attention}\label{pre:disentangle}

Unlike attention introduced in Section~\ref{pre:transformer}, disentangled attention uses relative position information directly when computing the attention scores~\cite{He:deberta}. 
The rationale for disentangled attention comes from the observation that relative distances are more important than absolute positions when calculating the attention score between tokens. In disentangled attention, the self-attention score for input $x = [x_1; \dots; x_n] \in \mathbf{R}^{n \times d}$ is calculated as:
\begin{align}\label{equation:disentangle}
    \begin{split}
    \alpha_{ij} =  Q(x_i)K(x_j)^T + Q(x_i)\hat{K}(\mathbf{p}_{\delta(i, j)})^T \\+ \hat{Q}(\mathbf{p}_{\delta(j, i)})K(x_j)^T
    \end{split}
\end{align}
where $\delta(i, j)$ refers to the relative distance between $x_i$ and $x_j$, $\mathbf{p} \in \mathbf{R}^{p \times d}$ is the relative position embedding where $p$ is the maximum relative length, and $Q, K, \hat{Q}, \hat{K} \in \mathbf{R}^{d \times d}$ are the projection matrices. While the first term remains the same as in vanilla self-attention, the second and third term calculates the relationship between $x_i$ and $x_j$ utilizing their relative distances. The output vector is calculated as:
\begin{align}\label{equation:disentangleagg}
    o_i =  \sum_j{\sigma(\frac{\alpha_{ij}}{\sqrt{3d}})V_j}
\end{align}
where $\sqrt{3d}$ comes from the three terms in Equation~\ref{equation:disentangle}. 

Tang~et.~al.~\cite{Tang:ast-trans} adopted a similar variant of this, using different output aggregation from Equation~\ref{equation:disentangleagg} (introduced later). They use two types of relative distances, parent-child and sibling distances, each of which is defined by two matices $P, S$. Parent-child relation matrix $P$ is defined as:

\begin{align}
    P_{ij}=\begin{cases}
    p_{ij}, & \text{ if i is ancestor of j},\\
    -p_{ij}, & \text{ if i is descendant of j} \\
    0, & \text{ otherwise }
    \end{cases}
\end{align}
\normalsize

where $p_{ij}$ is the shortest distance between node $i$ and $j$. Similarly, the sibling relation matrix $S$ is defined as,

\begin{align}
    S_{ij}=\begin{cases}
    s_{ij}, & \text{ if i, j are sibling and } p(i) < p(j),\\
    -s_{ij}, & \text{ if i, j are sibling and } p(i) > p(j) , \\
    0 & \text{ otherwise}
    \end{cases}
\end{align}
\normalsize
where $s_{ij}$ is the horizontal distance between node $i$ and $j$, and $p(i)$ is the child index of node $i$. Using these two matrices, $\delta(i, j)$ of each head are set to either $P_{ij}$ or $S_{ij}$ for multi-head disentangled attention. Attention scores whose corresponding distance is set to $0$ are masked, allowing attention calculation only for predefined node relationships. 




\subsection{Stochastic Block Model Attention}
Stochastic Block Model (SBM) is a graph generation model that generates node clusters. SBM is defined by three parameters: 1) $N$: the number of nodes, 2) $C_1, C_2 \dots C_k$: $k$ cluster of nodes, and 3) a symmetric matrix $S \in \mathbf{R}^{k \times k}$: $0 \le S_{ij} \le 1$ where $S_{i, j}$ denotes probability of a node in cluster $i$ connected to node in cluster $j$. 

Based on this, \cite{Cho:sbmattention} proposed SBM attention, which learns the node assignments and cluster relationship $S$. Using the learned matrices, we can calculate the probability of two nodes being connected. Such probabilities are then used dynamically to generate attention masks, to be used for self-attention. SBM attention is defined as follows.

A learnable vector is assinged for each of the $k$ clusters, yielding $C \in \mathbf{R}^{k \times d}$. Assuming that we have embedded an AST $x \in \mathbf{R}^{N \times d}$, we first calculate $\hat{Q}, \hat{K} \in \mathbf{R}^{N \times k}$.
\begin{align}
    \hat{Q} = QC^T,
    \hat{K} = KC^T,
\end{align}
where $Q, K \in \mathbf{R}^{N \times d}$ are the query and key vectors computed in self-attention. $\hat{Q}$ and $\hat{K}$ represent the probability of each node (i.e., rows) belonging to each cluster (i.e., columns). Using $C$, we compute $S = CC^T$, where entry $S_{ij}$ is the probability a node in cluster $i$ is connected to a node in cluster $j$. The probability matrix $P$ is calculated as $P = \hat{Q}S\hat{K}^T \in \mathbf{R}^{N \times N}$ where $P_{ij}$ calculates the expectation probability nodes $i, j$ are connected given $\hat{Q}, \hat{K}$ and $S$. Once we have $P$, Bernoulli sampling is used to decide whether to mask the entry $i, j$ for the attention score. Since the sampling process is a discrete operation, Straight-Through Estimator (STE) is used to let gradients flow. Specifically, the gradient regarding matrix $P$ is calculated as follows:

\begin{align}
   \frac{\partial{\emph{L}}}{\partial{P_{ij}}} \coloneqq  \frac{\partial{\emph{L}}}{\partial{M_{ij}}} = \begin{cases}
       \frac{\partial{\emph{L}}}{\partial{A_{ij}}} \cdot \frac{Q_iK_j^T}{\sqrt{d}} & \text{if } M_{ij} = 1 \\
       0 & \text{otherwise} \\
   \end{cases}
\end{align}
where $M$ is the sampled mask and $A \coloneqq M \odot \frac{QK^T}{\sqrt{d}}$. Therefore, while the mask is sampled from a Bernoulli distribution, the gradient flows as if we used matrix $P$ as masks and performed $P \odot \frac{QK^T}{\sqrt{d}}$. In practice, a sparsity regularization constant is used to penalize dense masks. For multi-head attention, different cluster embeddings are used so that each head learns separate attention masks.

\begin{python}[caption={Python functions},captionpos=b, label=listing:code1]
def config_option_show(context, data_dict):
    return {'success': False}
def test_abort():
    abort('Test')
\end{python}

\begin{figure}[h]
\centering
\subfigure[AST - config\_option\_show\label{fig:ast1}]{\includegraphics[angle=0, width=4cm]{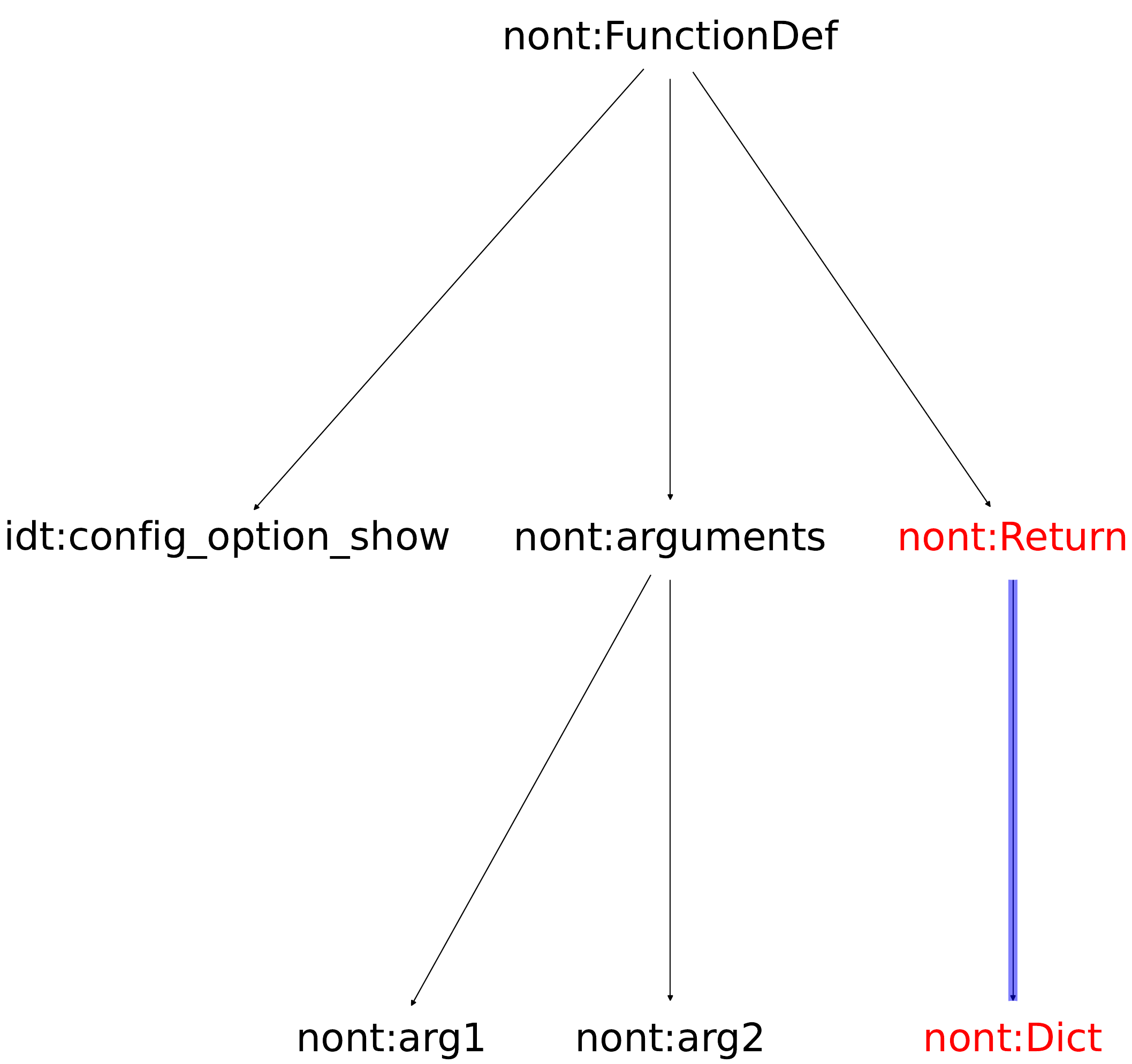}}
\subfigure[AST - test\_abort\label{fig:ast2}]{\includegraphics[angle=0, width=4cm]{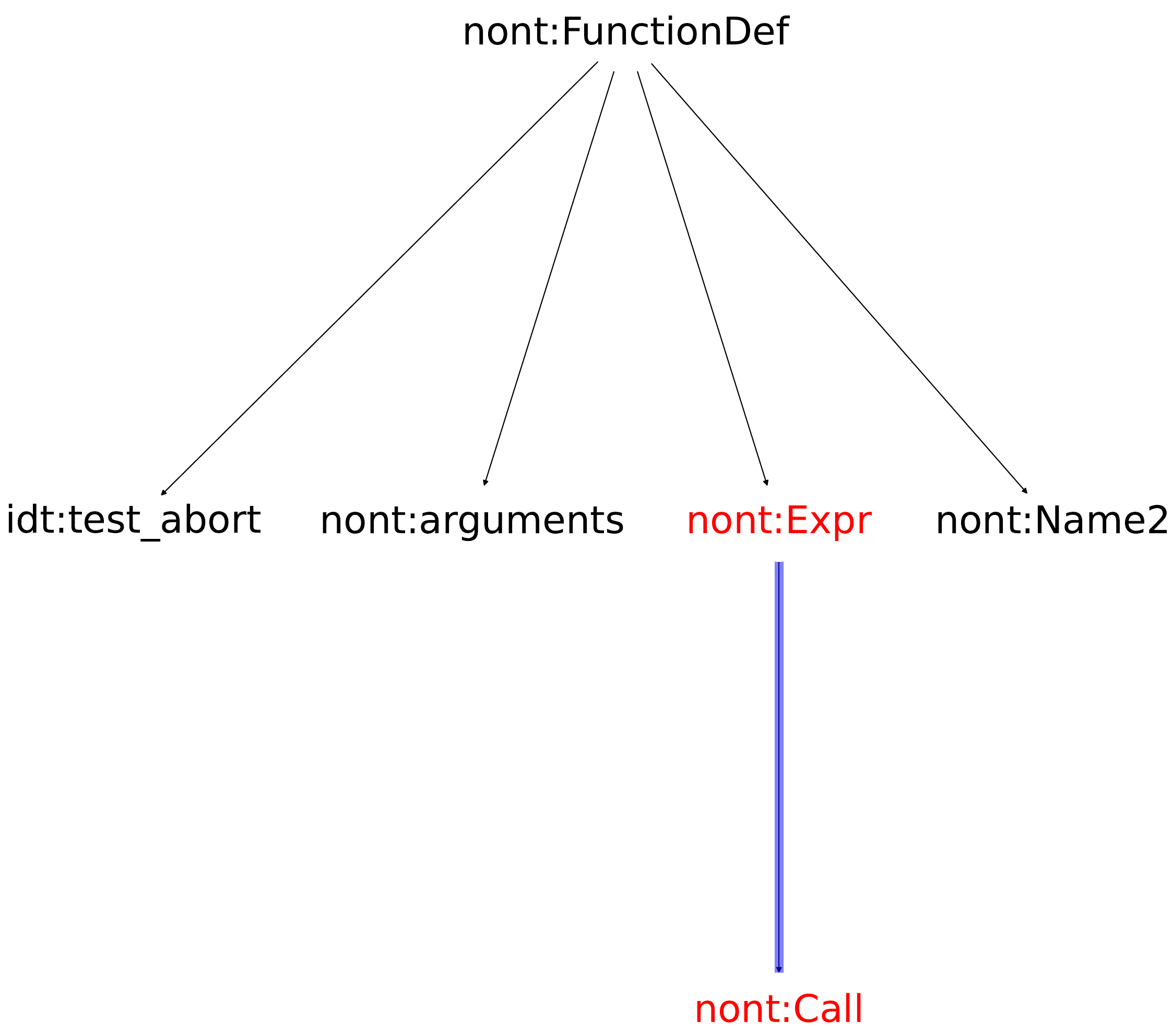}}

\caption{Program ASTs\label{fig:asts}}
\end{figure}
\section{Code Structure Aware Transformer}
\label{sec:csatrans}


\begin{figure*}[h]
\centering
\includegraphics[angle=0, width=15cm,]{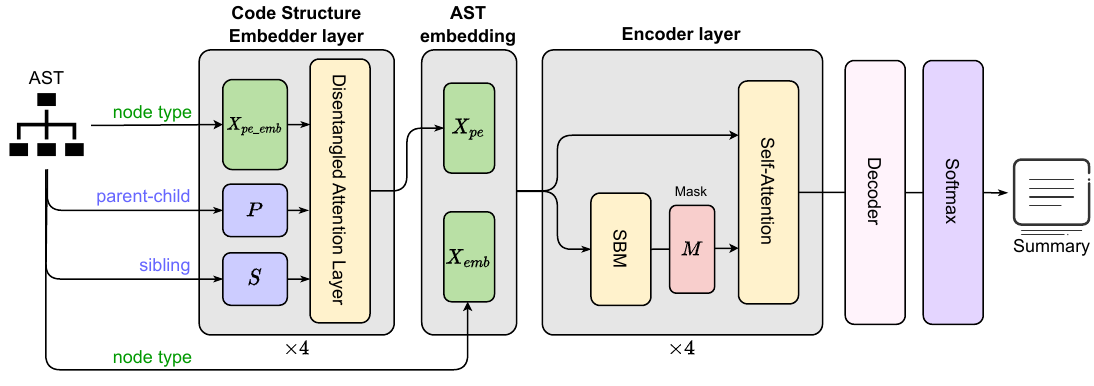}
\caption{CSA-Trans Architecture\label{fig:architecture}}
\end{figure*}

\subsection{Requirements for AST PE}\label{method:condition}

Let us consider the characteristics that an ideal PE for AST should have. First, we focus on its capacity to encode contextual information. We will use the two Python functions in Listing~\ref{listing:code1}, and their partial ASTs in Figure~\ref{fig:asts} as an example.
For each AST, consider node relationships; (\emph{Return}, \emph{Dict}) in Figure~\ref{fig:ast1} and (\emph{Expr}, \emph{Call}) in Figure~\ref{fig:ast2}. Note that these nodes for each AST are located in the same positions. Although their relationships (i.e., the blue edges) in the AST structure are the same, the semantics of these relationship differ. The first relation denotes that the return value of the function is of the dictionary type, whereas the second shows that the expression is a method call. If we are to only consider AST positions for PE, we would give the same PE for nodes (\emph{Expr}, \emph{Return}) as well as nodes (\emph{Dict}, \emph{Call}). Such PEs would not effectively communicate node relationships to the model. We argue that types of the target node, as well as its context (i.e., types of surrounding nodes as well as their relationships with the target node) should be incorporated into PEs. 

Second, we focus on the fact that ASTs are graphs. \emph{Permutation equivariance} is an important characteristic most GNN layers hold. Since nodes in graphs do not have fixed ordering, permuting node sequence should not affect how features of each node are calculated, including its PE. Accordingly, an ideal CSA-PE should be permutation equivariant; each node should get the same positional encoding regardless of how we permute the input node sequence.

Finally, an ideal PE for AST should generalize well to unseen AST structures. 


\subsection{Code Structure Embedder}
\label{method:pegen}
Higher performance of AST-Trans compared to transformer based models with global receptive field shows the gain from calculating attention between only the relevant nodes. Yet, the limitations are clear. Restricted attention calculation may hinder model from learning deeper node relationship and thus better summary. To overcome the problem but to take only the benefits from restricted disentangled attention, CSE performs attention mechanism similar to AST-Trans. Nevertheless, we use them as PE purpose only rather than using the output vectors to directly generate code summary.

Given an AST and max relative distance $p$, we first embed AST $X_{pe\_emb} \in \mathbf{R}^{N \times d_{pe}}$ using each node type. We also define two embeddings for each relation matrices (aforementioned parent-child and sibling) relations, $\mathbf{P}_{pe}, \mathbf{S}_{pe} \in \mathbf{R}^{p \times d_{pe}}$ ($\mathbf{p}$ in Equation~\ref{equation:disentangle}). Using $\mathbf{P}_{pe}, \mathbf{S}_{pe}$ and two relation matrices $P, S \in \mathbf{R}^{N \times N}$ of the AST ($\delta(\cdot)$ in Equation~\ref{equation:disentangleagg}), we get $X_{pe} \in \mathbf{R}^{N \times d_{pe}}$ of each node using attention introduced in Section~\ref{pre:disentangle}.

By incorporating relative information, each node will get different embedding depending on their node type \emph{and} nodes around them. Further, permutation equivariance of Multilayer Perceptron (MLP) and self-attention (permutation of node sequence would lead to change of output vectors in same permuted order.) let CSA-PE inherit the property. Finally since CSE uses two simple relationships which can be generalized, it does not suffer from out of distribution AST structures unlike Triplet PE.

CSE process differs from AST-Trans in attention aggregation. The aggregation for AST-Trans is shown in Equation~\ref{equation:asttransagg}. 

\begin{align}\label{equation:asttransagg}
    o_i =  \sum_j^{j \in \{j|\delta(i,j)>0\}}{\sigma(\frac{\alpha_{ij}}{\sqrt{3d}})(V_j + \hat{V}(\mathbf{p}_{\delta(i,j)}))}
\end{align}

It uses additional weight $\hat{V} \in \mathbf{R}^{d \times d}$ to give extra relative position information to the value vectors. While such aggregation may profit the model to learn richer relative node information, it is time consuming since the operation cannot be done in simple addition. To alleviate the time complexity, AST-Trans uses sparse tensor and additional constraint; a node is updated using only partial nodes among predefined relevant node set. ($j \in \{j|\delta(i,j)>0\}$) Yet, CSE uses simple summation based aggregation as in Equation~\ref{equation:disentangleagg}. In such way, CSE can be more efficient since the aggregation can be done via simple addition. In turn, it hands over the job of learning richer node relationship to the encoder. Also, while AST-Trans sets max relative distance to 5, 10 for each parent-child and sibling relationship, our simple aggregation allows to enlarge the max relative distance with small additional memory. For AST-Trans, the aggregation approach results in a memory footprint of $O(pNd)$, where $p, N$ and $d$ are each maximum relative distance, number of tokens, and the hidden dimension. In contrast, our aggregation strategy maintains a more efficient memory footprint of $O(pN)$. To ensure that the memory footprint remains within an acceptable range of $O(N^2)$, i.e., that of attention, we have opted to set the maximum relative distance equal to $N$. Therefore, we set max relative distance to $150$. Given that the maximum distance between nodes for tree with $N$ nodes is $N-1$, our choice includes all node pairs.

\subsection{Encoder}\label{method:encoder}
Recall that CSA-PE is \emph{context-aware}; node PE has been updated considering its sibling and ancestor, descendant nodes. Even if two nodes have the same node type, their embeddings are differentiated in their PE. Therefore, we put aside the hard-coded mask used for filtering relevant nodes and let SBM attention learn to select the relevant nodes.

Given $X_{pe\_emb} \in \mathbf{R}^{N \times d_{pe}}$. We embed each AST node using node type, yielding $X_{emb} \in \mathbf{R}^{N \times d_{emb}}$. We then concatenate CSA-PE and node embedding, $X \in \mathbf{R}^{N \times (d_{pe} + d_{emb})}$. Embedding $X$ is fed into the encoder, whose attention scores are calculated via SBM attention.

The encoder infers relations between nodes from PEs, rather than from hard-coded as in AST-Trans~\cite{Tang:ast-trans} and SG-Trans~\cite{Gao:sg-trans}. Consider the \emph{Dict} node in Figure~\ref{fig:ast1}. It is neither in parent nor sibling relation with \emph{arg1}, if we were to use attention restricted by parent-child and sibling relation, it is not possible for the \emph{Dict} node to be directly affected by the node \emph{arg1}. On the other hand, if SBM attention regards this pair of nodes to be relevant, it can decide to not mask out the pair and use them to update embedding of each other.

\section{Evaluations}
\label{sec:experiments}

\subsection{RQ1. How good is CSA-Trans for code summarization?}

\subsubsection{Dataset}
We evaluate code summarization performance on public Java~\cite{Hu:API+Code} and Python~\cite{Wan:RLHybrid} datasets, following SG-Trans~\cite{Gao:sg-trans}. The Java dataset contains 87,138 Java method and comment pairs from 9,714 GitHub repositories; the Python dataset consists of 92,545 functions and their docstrings. We use the same dataset split used in SG-Trans~\cite{Gao:sg-trans}. Dataset statistics can be found in Table~\ref{tbl:datastat}.
\begin{table}[h]
\centering
\caption{Dataset Statistics}
\label{tbl:datastat}
\begin{tabular}{lrrr}
\toprule
Language & Train & Validation & Test \\ \midrule
Java & 69,708 & 8,714 & 8,714 \\ 
Python & 55,538 & 18,505 & 18,502  \\
\bottomrule
\end{tabular}
\end{table}

\subsubsection{Preprocessing}
We generate our AST using \emph{tree-sitter}\footnote{https://tree-sitter.github.io/tree-sitter/}. We use the maximum source length of 150 nodes and the maximum output length of 50 summary words, to match settings with SG-Trans. For each identifier in AST, we apply \emph{CamelCase} and \emph{snake\_case} tokenization, after which subtokens are connected vertically in the AST.

\subsubsection{Evaluation Metrics}
We use three evaluation metrics. \emph{BLEU}~\cite{Papineni:bleu} is n-gram based similarity matching algorithm used in various natural language and software engineering literature~\cite{Jiang:bleuusage, Zhang:bleuusage2}. \emph{METEOR}~\cite{Banerjee:meteor} calculates the ability model output can capture the reference text by aligning reference and generated sentence. \emph{ROUGE-L}~\cite{Lin:rouge} calculates how much the generated text recovers the reference text based on the Longest Common Subsequence (LCS) between generated output and reference.


\begin{table*}[t]
\centering
\caption{Code Summarization Performance}
\label{tbl:codesum}
\begin{tabular}{lcrrr|lrrr}
\toprule
Language & Representation & \multicolumn{3}{c}{Java} & \multicolumn{3}{c}{Python}  \\
                          & & BLEU-4 & METEOR & ROUGE-L & BLEU-4 & METEOR & ROUGE-L \\ \midrule
CODE-NN~\cite{Iyer:CodeNN} & Code & 27.60 & 12.61 & 41.10 & 17.36 & 09.29 & 37.81 \\
Tree2Seq~\cite{Eriguchi:Tree2Seq} & Code & 37.88 & 22.55 & 51.50 & 20.07 & 08.96 & 35.64 \\
API+Code~\cite{Hu:API+Code} & Code & 41.31 & 23.73 & 52.25 & 15.36 & 08.57 & 33.65 \\
Dual Model~\cite{Wei:DualMode} & Code & 42.39 & 25.77 & 53.61 & 21.80 & 11.14 & 39.45 \\
Vanilla Transformer~\cite{Vaswani:transformer} & Code & 44.20 & 26.83 & 53.45 & 31.34 & 18.92 & 44.39 \\
NeuralCodeSum~\cite{20Ahmad:NeuralCodeSum} & Code & 45.15 & 27.46 & 54.84 & 32.19 & 19.96 & 46.32 \\
SG-Trans~\cite{Gao:sg-trans} & Code & 45.48 & 27.50 & 55.44 & 33.02 & 20.45 & 46.93 \\
\midrule
Code2seq~\cite{code2seq:19Alon} & AST (Path) & 12.19 & 08.83 & 25.61 & 18.69 & 13.81 & 34.51 \\
DeepCom~\cite{Hu:badcomment} & Code + AST & 39.75 & 23.06 & 52.67 & 20.78 & 09.98 & 37.35 \\
Transformer+GNN~\cite{Choi:Graph+Transformer} & AST & 45.49 & 27.17 & 54.82 & 32.82 & 20.12 & 46.81 \\ 
AST-Trans~\cite{Tang:ast-trans} & AST & 45.53 & 28.89 & 55.30 & - & - & - \\
\midrule
RL + Hybrid2Seq~\cite{Wan:RLHybrid} & Code + AST & 38.22 & 22.75 & 51.91 & 19.28 & 09.75 & 39.34 \\
GREAT~\cite{great:20Hellendoorn} & Code + AST & 44.97 & 27.15 & 54.42 & 32.11 & 19.75 & 46.01 \\
\midrule
CodeT5-small~\cite{Wang:Codet5} & Code & 43.16 & 28.31 & 53.93 & 33.45 & 21.16 & 48.94 \\
\midrule
Ours (CSA-Trans) & AST & \textbf{46.02} & \textbf{29.39} & \textbf{56.10} & \textbf{35.81} & \textbf{21.26} & \textbf{49.16} \\ \bottomrule
\end{tabular}
\end{table*}

\subsubsection{Baselines}
We use the following 15 baselines code summarization models, we categorize baseline models by the modality each model uses to embed AST.

\noindent\textbf{Source Code as Language}

\emph{CODE-NN}~\cite{Iyer:CodeNN} and \emph{Tree2Seq}~\cite{Eriguchi:Tree2Seq} each uses LSTM and Tree-LSTM architecture to generate code summary. \emph{API+CODE}~\cite{Hu:API+Code} uses API call sequences with multiple encoders. \emph{Dual Model}~\cite{Wei:DualMode} uses dual learning and train model on both code summarization and generation. \emph{Vanilla Transformer}~\cite{Vaswani:transformer} uses Transformer architecture. \emph{NeuralCodeSum}~\cite{20Ahmad:NeuralCodeSum} uses relative position encoding and copy attention. \emph{SG-Trans}~\cite{Gao:sg-trans} uses different token relationships and use various attention masks along with copy attention. CodeT5~\cite{Wang:Codet5} is a pretrained transformer model trained using code-related pretraining task such as masked identifier prediction. It is pretrained on source codes of different languages including 3,158,313 code snippets with their natural langauge comments. In order to match the number of parameters, we compare the performance with \emph{CodeT5-small}.

\noindent\textbf{Source Code as AST} 

\emph{Code2Seq}~\cite{code2seq:19Alon} represents source code with AST paths where each path is embeded using Bi-directional LSTM. \emph{DeepCom}~\cite{Hu:badcomment} first utilizes the Structure-based Traversal (SBT) and use LSTM to encode the linearized AST. \emph{Transformer + GNN}~\cite{Choi:Graph+Transformer} encodes the AST with GNN, which is then passed onto Transformer architecture. \emph{AST-Trans}~\cite{Tang:ast-trans} uses two kinds of node relative relationship with disentangled attention.

\noindent\textbf{Hybrid approaches}

\emph{RL+Hybrid2Seq}~\cite{Wan:RLHybrid} uses code and AST information with reinforcement learning. \emph{GREAT}~\cite{great:20Hellendoorn} uses code tokens with various relational information obtained from AST. \emph{\textsc{CodeScribe}}~\cite{guo:codescribe} uses both AST and source code with triplet node positions, where both embeddings are incorporated in building AST embedding. Pointer generator network is used to pay attention to both code and AST.

Performance of the baseline models except for AST-Trans, SG-Trans, CodeT5, and \textsc{CodeScribe} are taken from the literature~\cite{Gao:sg-trans}. Since AST-Trans was evaluated on tokenized summary\footnote{Other baselines use un-tokenized summary.} and used different settings, we replicated AST-Trans using un-tokenized summary and matched max input/output (150/50) length. We also increase the encoder/decoder dimension to $512$ to match model size with other baselines. Since the Python dataset of AST-Trans is different, we also train CSA-Trans using their data split and compare the results separately.\footnote{Their Python AST preprocessing was not applicable to dataset used by SG-Trans and ours.} Similarly, since \textsc{CodeScribe} uses different preprocessed summary, we retrain CSA-Trans with their summary and compare the results separately. We set beam width of \textsc{CodeScribe} to 1 to match settings. We note that \textsc{CodeScribe} uses maximum of (317, 535) code tokens and AST nodes for Python and (285, 361) for Java, which are more than 150 used by CSA-Trans. However, the available implementation does not easily allow changing these hyperparameters, hindering fair comparison. Therefore for Java comparison with \textsc{CodeScribe}, we use maximum 300 nodes to train CSA-Trans. We include specific parameter details in Section~\ref{section:parameter} where CSA-Trans uses smaller number of parameters compared to replicated AST-Trans and SG-Trans.
It is important to acknowledge that, being a pretrained model, CodeT5 depends on a pretrained tokenizer and its corresponding embedding. Therefore, setting the maximum output length to 50 would truncate the summary leading to incomparable results. We therefore increase the maximum output length to 93/429 for Java/Python dataset, in which CodeT5 is then able to generate the full summary we use in our evaluation. We finetune CodeT5 for 60/30 epochs for each Java and Python dataset with learning rate of $10^{-3}$ and dropout rate of 0.2. The decision to finetune longer for the Java dataset was prompted by the higher levels of observed loss in comparison to the Python dataset.




\subsubsection{Parameter Settings}
\label{section:parameter}
We set $d_{pe} = 256$ and $d_{emb} = 256$ for Python and $d_{pe} = 128$ and $d_{emb} = 640$ for Java. For Java, we find that decreasing the PE dimension and increasing encoder dimension leads to better performance. The decoder hidden dimension is fixed to $512$. CSE, encoder, and decoder of CSA-Trans all have four layers. We note that the resulting number of trainable parameters is smaller than both AST-Trans and SG-Trans that we have replicated. We report the number of parameter for CSA-Trans, AST-Trans, SG-Trans, and CodeT5 in Table~\ref{tbl:paramcomp}. We set the learning rate as $10^{-4}$ and use the AdamW optimizer~\cite{Loshchilov:adamw}. For SBM attention, we set the number of clusters to $10$ and the sparsity regularization constant to $10^{-2}$. We set maximum vocabulary size to 10,000 and 20,000 for AST node types and target summary respectively.

\begin{table}[h]
\centering
\caption{Number of parameters}
\label{tbl:paramcomp}
\begin{tabular}{l|rr}
\toprule
Models & Java & Python \\ \midrule
CSA-Trans & 72198176 & 60344864\\ \midrule
AST-Trans & 73190978 & 80518147\\ \midrule
SG-Trans & 91635675 & 100654897\\ \midrule
CodeT5 & 60492288 & 60492288 \\ \bottomrule
\end{tabular}
\end{table}

\subsubsection{Result}
\label{sec:rq1result}

Table~\ref{tbl:codesum} shows the code summarization performance of CSA-Trans and 14 baselines. CSA-Trans shows the best performance, achieving the highest score for all studied metrics. For Java, we get 0.49\%p, 0.5\%p, and 0.8\%p increase in BLEU-4, METEOR, ROUGE-L, respectively, when compared to the second-best performing model for Java, AST-Trans. For Python, we get 2.36\%p, 0.10\%p, and 0.22\%p increase in BLEU-4, METEOR, and ROUGE-L repectively compared to CodeT5. Although we finetuned CodeT5 for 60/30 epochs which are longer than the 15 epoch used in the code summarization evaluation of CodeT5~\cite{Wang:Codet5}, the performance gap between CodeT5 and CSA-Trans shows how a dedicated model can be of value for a specific downstream task.


Among baselines, those with Transformer architecture that are pretrained or use either the AST for representation or the structural information from AST show stronger performances. This shows that structural information does contribute to better performance in code summarization. Table~\ref{tbl:asttrans} shows the results of the separate comparison of CSA-Trans with AST-Trans on Python dataset. CSA-Trans achieves 0.91\%p, 1.46\%p, and 1.72\%p higher scores than AST-Trans on studied metrics. The higher scores achieved by CSA-Trans show the gain from utilizing global receptive field enabled by SBM attention.


\begin{table}[h]
\centering
\caption{Comparison with AST-Trans for Python}
\label{tbl:asttrans}
\begin{tabular}{l|rrr}
\toprule
Metrics & BLEU-4 & METEOR & ROUGE-L \\ \midrule
AST-Trans & 34.95 & 20.05 & 47.54  \\ \midrule
Ours (CSA-Trans) & 35.86 & 21.51 & 49.26 \\ \bottomrule
\end{tabular}
\end{table}

\begin{table}[]
\caption{Comparison with CODESCRIBE}
\label{tbl:codescribe}
\begin{tabular}{lcrrr}
\toprule
 & & BLEU-4 & METEOR & ROUGE-L \\ \hline
\multicolumn{1}{l|}{\multirow{2}{*}{Java}} & \multicolumn{1}{l|}{CODESCRIBE} & 46.70 & 31.44 & 58.38 \\ \cline{2-2}
\multicolumn{1}{l|}{} & \multicolumn{1}{l|}{CSA-Trans} & 46.79 & 30.96 & 58.38 \\ \hline
\multicolumn{1}{l|}{\multirow{2}{*}{Python}} & \multicolumn{1}{l|}{CODESCRIBE} & 32.65 & 22.42 & 48.71 \\ \cline{2-2}
\multicolumn{1}{l|}{} & \multicolumn{1}{l|}{CSA-Trans} & 35.79 & 22.97 & 50.69 \\
\bottomrule
\end{tabular}
\end{table}

Similarly, Table~\ref{tbl:codescribe} shows the comparison of CSA-Trans with \textsc{CodeScribe}. For Java, CSA-Trans shows mixed results by achieving 0.09\%p increase in BLEU-4 while losing by 0.48\%p on METEOR. However, we also observe strong performance of CSA-Trans for Python, achieving 3.11\%p, 0.46\%p, and 1.92\%p increase. The large number of code tokens and AST nodes used by \textsc{CodeScribe} may have contributed to the higher performance.


\subsection{RQ2: Does CSA-PE provide better PE?}
In this section, we investigate how much CSA-PE contributes to the code summarization performance.

\subsubsection{Intermediate Node Prediction}

 The objective of PE is to provide positional information so that the model can distinguish and find relationships between different nodes. Therefore, a good PE should be able to infer the relationship between two nodes directly from the embedding without external bias. To evaluate such capabilities, we have devised a synthetic experiment Intermediate Node Prediction (INP). The goal of the task is to let Multilayer Perceptron (MLP) predict the sequence of intermediate nodes given two end node PEs. Our aim is to empower each node's PE to infer relational information via straightforward operations. This design choice enables our summarization model to infer node relationships through attention modules and feedforward networks. Consequently, we have intentionally chosen a simple and straightforward technique, MLP, to assess the capability of PE itself. If the MLP is able to learn to predict intermediate nodes with high accuracy, the PE can be seen to have richer information about the node context and the summarization model using the PE is expected to learn better node relationships via attention mechanism.

 Specifically, k-INP aims to predict k nodes between two end nodes. Figure~\ref{fig:synthetic} shows an example of 3-INP instance, where MLP needs to predict the sequence \emph{Block}, \emph{While}, and \emph{Expr} given the positional encoding of \emph{Function Def} and \emph{Method Invocation}. We perform 1, 3, 5-INP tasks, whereas we run each task 10 times to get reliable results.

\tikzstyle{wc} = [circle, rounded corners, minimum width=1cm, minimum height=1.1cm,text centered, draw=black, fill=white, font=\fontsize{5pt}{5pt}\selectfont]
\tikzstyle{yc} = [circle, rounded corners, minimum width=1cm, minimum height=1.1cm,text centered, draw=black, fill=yellow, font=\fontsize{5pt}{5pt}\selectfont]
\tikzstyle{arrow} = [thick,->,>=stealth]
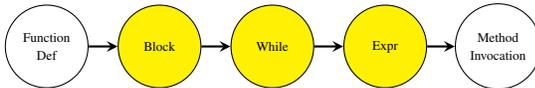
\begin{figure}[ht]
\centering
  \begin{tikzpicture}[node distance=1.5cm, auto]
    \node (comp1) [wc, xshift=2cm] {\shortstack[alignment]{Function\\Def}};
    \node (comp2) [yc, right of=comp1] {  Block  };
    \node (comp3) [yc, right of=comp2] { While };
    \node (comp4) [yc, right of=comp3] { Expr };
    \node (comp5) [wc, right of=comp4] { \shortstack[alignment]{Method \\Invocation}};
    
    \draw [arrow] (comp1) -- (comp2);
    \draw [arrow] (comp2) -- (comp3);
    \draw [arrow] (comp3) -- (comp4);
    \draw [arrow] (comp4) -- (comp5);
  \end{tikzpicture}
\caption{Intermediate Node Prediction: 3-INP prediction}
\label{fig:synthetic}
\end{figure}

\subsubsection{Dataset}
To construct the dataset for k-INP, we first calculate all node pairs of length k+1. From each AST, we extract at most 10 pairs to build the dataset. We split the datasets into train and test set with the ratio of 8:2. We report the dataset statistics in Table~\ref{tbl:inpdataset}. We use MLP model with two hidden layers for INP, using CSA-PE as well as baselines that include Sequential PE, Tree PE, Triplet PE, and Laplacian PE.


\begin{table}[h]
\centering
\caption{Number of Train/Test node sequences for k-INP}
\label{tbl:inpdataset}
\begin{tabular}{l|cc}
\toprule
k-INP & Java & Python \\ \midrule
k=1 & 69669 / 17418 & 148016 / 37004\\ \midrule
k=3 & 69591 / 17398 & 148012 / 37004\\ \midrule
k=5 & 69406 / 17352 & 147776 / 36944\\ \bottomrule
\end{tabular}
\end{table}

\begin{table}[t]
\centering
\caption{Intermediate Node Prediction Accuracy}
\label{tbl:synthetic}
\begin{tabular}{lrrr}
\toprule
Language & \multicolumn{3}{c}{Java} \\
                          & k=1 & k=3 & k=5 \\ \midrule
Sequential PE & 36.08 ± 0.73 & 13.41 ± 0.8 & 4.17 ± 0.53 \\
Tree PE & 42.14 ± 1.11 & 22.72 ± 4.08 & 16.1 ± 2.9 \\
Triplet PE & 34.19 ± 0.37 & 11.57 ± 2.57 & 2.67 ± 0.61 \\
Laplacian PE & 37.01 ± 1.76 & 11.04 ± 1.05 & 4.24 ± 1.01 \\ \midrule
CSA-PE & \textbf{78.04 ± 3.03} & \textbf{55.05 ± 2.81} & \textbf{35.48 ± 3.23} \\ \bottomrule
\toprule
Language & \multicolumn{3}{c}{Python} \\
 & k=1 & k=3 & k=5 \\ \midrule \\
Sequential PE & 34.16 ± 0.46 & 9.67 ± 0.33 & 2.03 ± 0.15 \\
Tree PE & 57.69 ± 2.48 & 37.77 ± 2.66 & 23.42 ± 2.22 \\
Triplet PE & 18.1 ± 0.18 & 0.38 ± 0.4 & 0.09 ± 0.07  \\
Laplacian PE & 37.89 ± 2.73 & 12.82 ± 1.06 & 3.55 ± 0.33 \\ \midrule
CSA-PE & \textbf{76.95 ± 5.05} & \textbf{54.15 ± 3.99} & \textbf{30.67 ± 3.46} \\ \bottomrule
\end{tabular}
\end{table}

\subsubsection{Result}

The result of INP is shown in Table~\ref{tbl:synthetic}. Notably, CSA-PE consistently yields the best result across all k-INP tasks. CSA-PE obtains significantly higher accuracies, having 35.90\%, 19.26\% better performance in 1-INP compared to second best performing Tree PE. It is worth mentioning that Tree PE achieves substantially higher scores than other baselines, underscoring that the structural aspect of AST can help neural networks infer the relationship of AST nodes. The tendency carries onto 3-INP and 5-INP tasks where CSA-PE still obtains meaningful higher scores compared to all other baselines. The result shows that the relationship between nodes can be well inferred using CSA-PE compared to other baseline PEs, leading to performance enhancements.


\begin{table}[]
\caption{Positional Encodings}
\label{tbl:pecompare}
\centering
\begin{tabular}{l|c|rrr}
\toprule
                        & PE scheme & BLEU-4 & METEOR & ROUGE-L \\ \hline
\multirow{5}{*}{Java}   & Sequential PE & 35.26  & 20.70  & 48.08   \\
                        & Tree PE       & 35.76  & 21.04  & 48.69   \\
                        & Triplet PE    & 35.52  & 20.92  & 48.42   \\
                        & Laplacian PE  & 34.82  & 20.01  & 46.85   \\
                        & CSA-PE        & \textbf{35.81}  & \textbf{21.26}  & \textbf{49.16}   \\ \hline
\multirow{5}{*}{Python} & Sequential PE & 45.49  & 28.86  & 55.06   \\
                        & Tree PE       & 45.50  & 28.65  & 54.98   \\
                        & Triplet PE    & 45.18  & 28.41  & 50.40   \\
                        & Laplacian PE  & 44.62  & 28.11  & 53.64   \\
                        & CSA-PE        & \textbf{46.02}  & \textbf{29.39}  & \textbf{56.10}   \\ \bottomrule
                        
\end{tabular}
\end{table}

\subsubsection{Ablation Study on PE} 

In order to see how different PEs directly affect the model performance, we train five variants of CSA-Trans with different PE schemes. Other than Sequential PE, we set same encoding dimension as CSA-PE and concatenate the resulting PE to AST embedding. Since Sequential PE is usually added (instead of concatenated) to the code embeddings, we enlarge the embedding dimension to match the encoder hidden dimension and add the PE to the embedding vector following the original Transformer~\cite{Vaswani:transformer}. 

The result is shown in Table~\ref{tbl:pecompare}. The best metrics are obtained by utilizing CSA-PE, outperforming baselines in every metrics. This aligns with our previous results on INP task showing how providing PE with contextual information can help model to generate better summary. Aside from CSA-PE, we observe Tree PE and Sequential PE perform better than Laplacian PE. The performance of Laplacian PE suggests that AST may have different characteristics from general graphs.


\begin{table}[t]
\caption{Model Ablation on SBM attention}
\label{tbl:sbmattn}
\begin{tabular}{l|c|rrr}
\toprule
 Dataset & Attention Method & BLEU-4 & METEOR & ROUGE-L \\ \hline
\multicolumn{1}{l|}{\multirow{2}{*}{Java}} & \multicolumn{1}{l|}{Vanilla-Attention} & 45.64 & 29.03 & 55.27 \\ \cline{2-2}
\multicolumn{1}{l|}{} & \multicolumn{1}{l|}{SBM-Attention} & 46.02 & 29.39 & 56.10 \\ \hline
\multicolumn{1}{l|}{\multirow{2}{*}{Python}} & \multicolumn{1}{l|}{Vanilla-Attention} & 35.38 & 20.69 & 48.32 \\ \cline{2-2}
\multicolumn{1}{l|}{} & \multicolumn{1}{l|}{SBM-Attention} & 35.81 & 21.26 & 49.16 \\
\bottomrule
\end{tabular}
\end{table}
\subsection{RQ3: What benefits does SBM attention bring?}
\subsubsection{Quantitative analysis on SBM attention}
We investigates the contribution dynamically learned SBM masks, both quantitatively and qualitatively. In order to see the effect of SBM attention, we train CSA-Trans with vanilla self-attention and compare the result. Table~\ref{tbl:sbmattn} contains the comparison between variants of CSA-Trans that use SBM and vanilla attention respectively. SBM attention outperforms the vanilla attention mechanism across all metrics and datasets. This shows \emph{learning} which AST nodes to attend rather than calculating relationship between all pair of nodes can contribute to higher summarization performance. 



\begin{figure}[h]
\subfigure[Java \label{fig:sbmmaskplotjava}]{\includegraphics[angle=0, width=4cm]{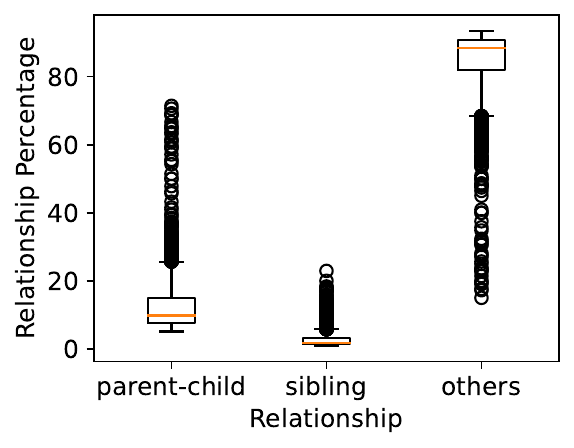}}
\subfigure[Python \label{fig:sbmattplot}]{\includegraphics[angle=0, width=4cm]{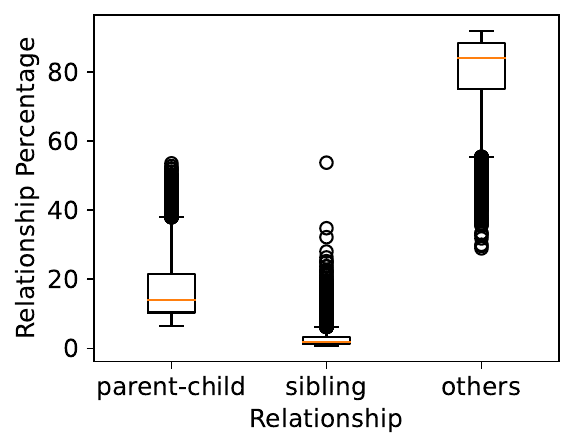}}
\caption{Relationships induced by SBM mask}
\label{fig:sbmmaskplot}
\end{figure}

We have also conducted an analysis of the relationships between node pairs for which the SBM mask enables attention calculation. Figure~\ref{fig:sbmmaskplot} shows the percentage of each relationship among all attention calculations of each ASTs on test split for each dataset. We observe about 80.35\% / 85.33\% of the node pairs fall into neither parent-child nor sibling relationship for Java and Python respectively. This implies that if we were to use the restricted attention within parent-child and sibling relationships, a substantial portion of node pairs selected by SBM attention would be masked. On top of that, the performance gap between AST-Trans and CSA-Trans illustrates how \emph{learning} which nodes to attend instead of using hardcoded masks can yield notable summarization performance enhancements.

\subsubsection{Qualitative analysis on SBM attention}
\begin{figure}[t]
\centering
\includegraphics[angle=0, width=4cm]{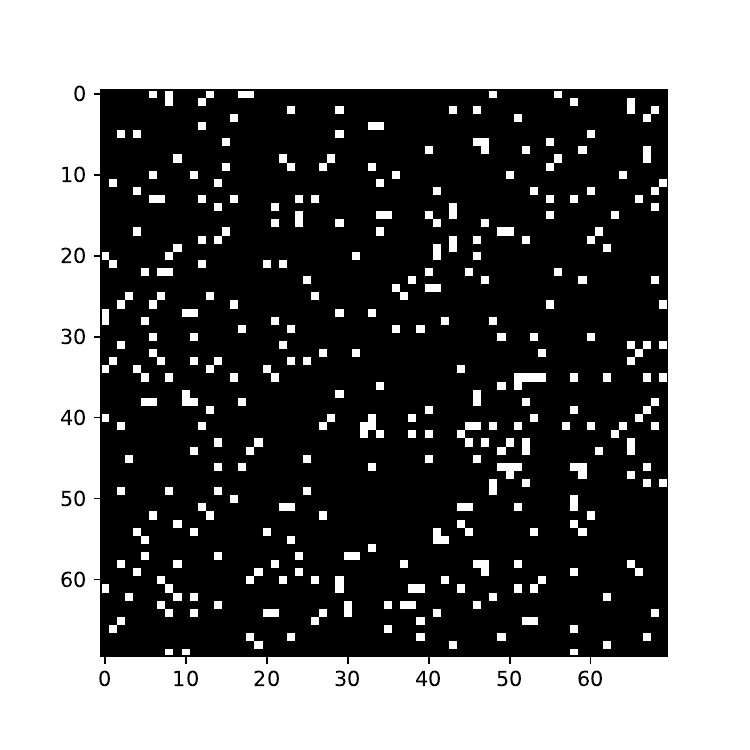}
\caption{Attention Masks generated by SBM attention.}
\label{fig:sbmmask}
\end{figure}

For qualitative analysis, we visualize the SBM attention mask and scores. Figure~\ref{fig:sbmmask} shows the aggregated masks generated from multi-head SBM attention. The selected node pairs are shown in white, where the attention coefficients in the black region are chosen to be masked to 0. We note that SBM masks lack locality, that is, nodes attends to other nodes that are far apart when the AST is linearized. We conjecture this is due to the structured nature of AST, where importance of node relationships does not always correlate to linearized node distances. 

 \begin{python}[caption={Python function},captionpos=b, label=listing:code2]
def upper(s):
  return s.upper()
\end{python}

\begin{figure}[h]
\centering
\includegraphics[angle=0, width=8cm]{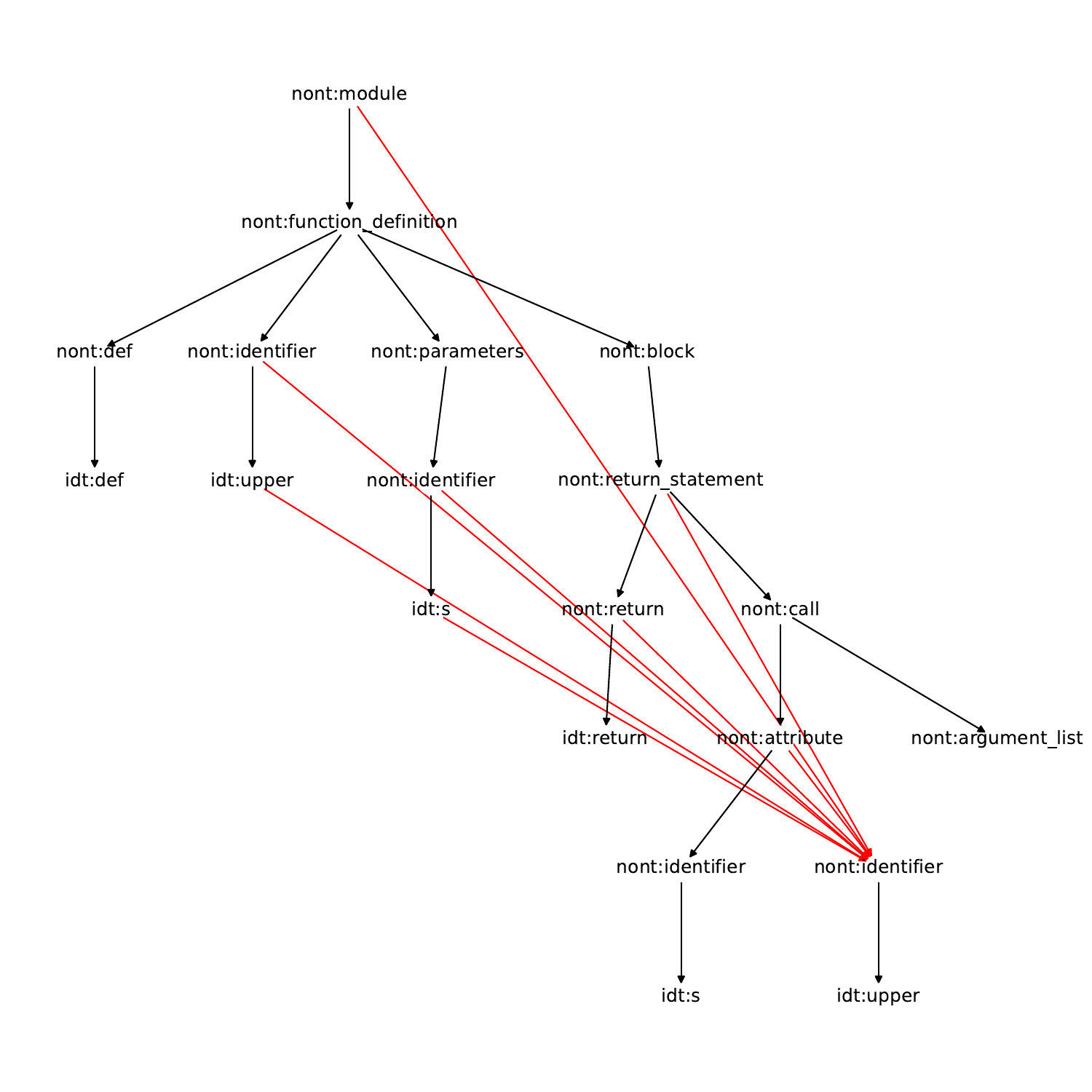}
\caption{Visualization of attention mask on AST.}
\label{fig:sbmmast}
\end{figure}

To further analyze, we visualize the SBM attention mask for a simple function in Listing~\ref{listing:code2}. The nodes which SBM calculates attention at least once with \emph{upper} are connected with the \emph{upper} node using red arrows in Figure~\ref{fig:sbmmast}. We observe that the attention is calculated with other identifier nodes such as \emph{upper} and \emph{s}, which are highly related to the target identifier in terms of code semantics. The SBM attention also calculated with the \emph{return} node, which defines the role of the target node. The example suggests that CSA-Trans is able to filter out less relevant nodes, and learn to aggregate only from the more relevant nodes by generating dynamic attention mask. It is notable to highlight that merely two out of the eight nodes employed by SBM attention for updating the \emph{identifier} node falls into to parent-child or sibling relationships. This observation highlights the significance of employing an attention mechanism that is not confined by such restrictions.



\begin{figure}[h]
\centering
\subfigure[SBM Attention Coefficient\label{fig:sbmcoef}]{\includegraphics[angle=0, width=6cm]{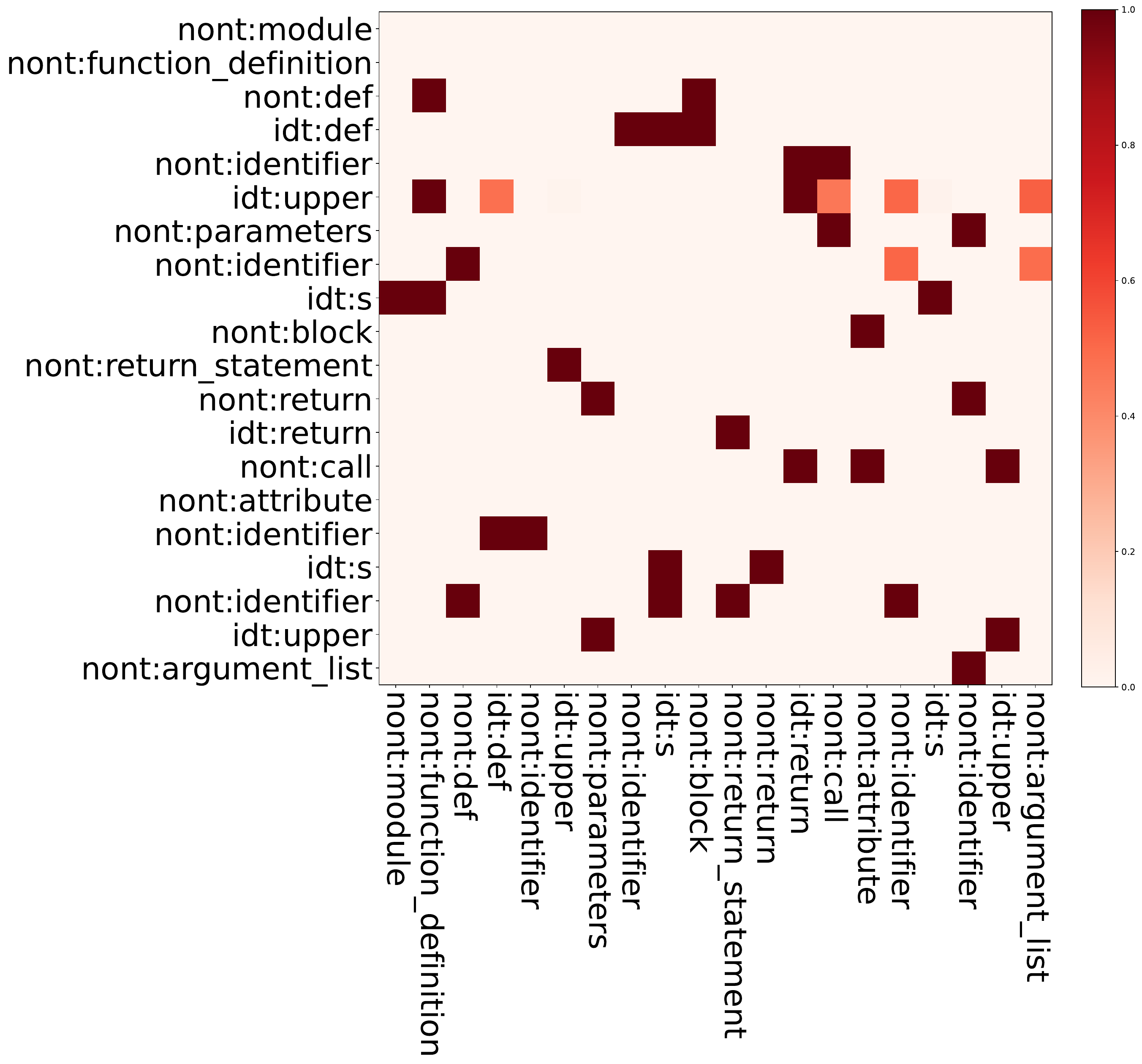}}

\subfigure[Vanilla Attention Coefficient\label{fig:fullcoef}]{\includegraphics[angle=0, width=6cm]{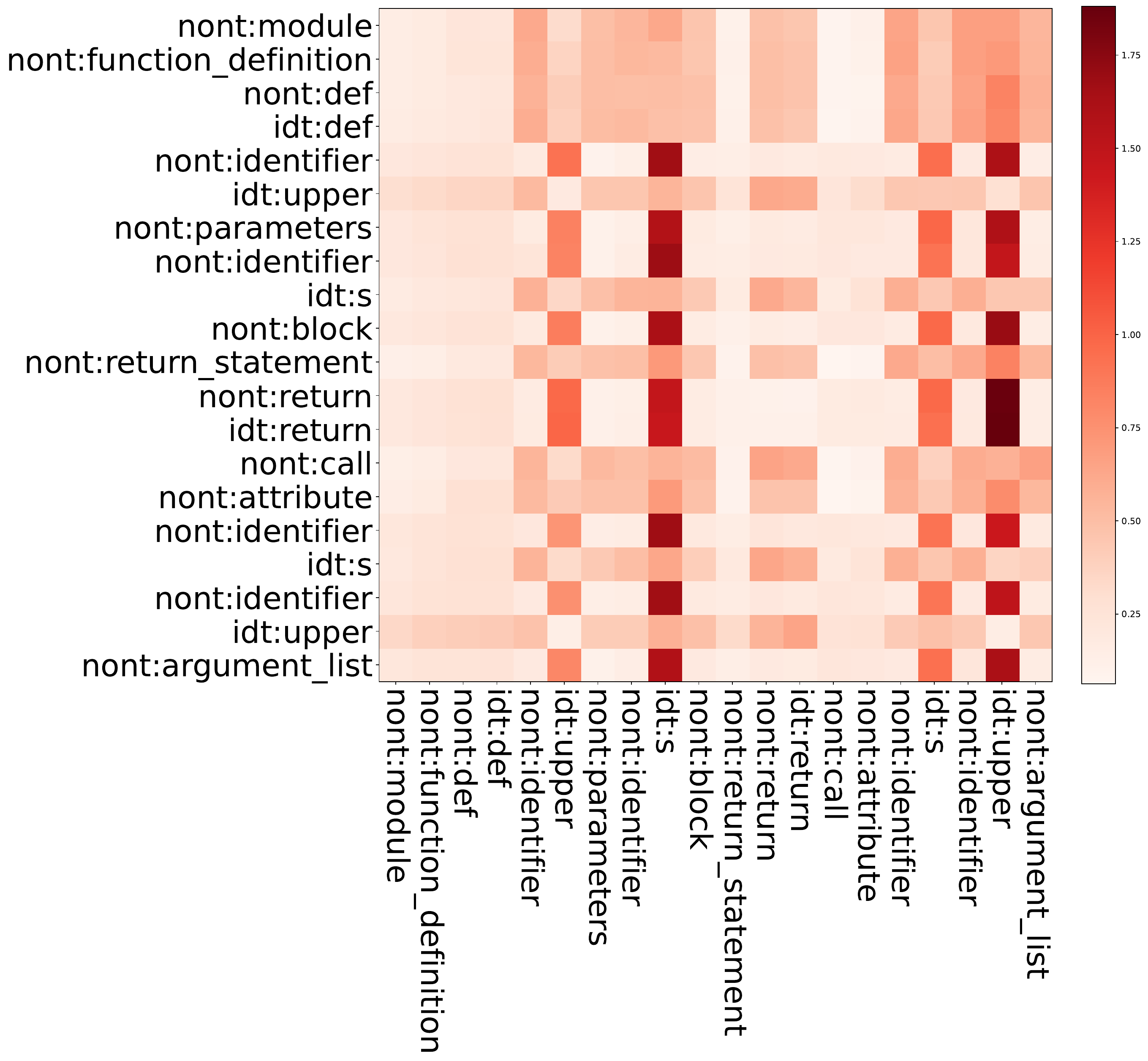}}

\caption{SBM and vanilla attention coefficients.}
\label{fig:sbmatt}
\end{figure}

We also visualize the attention scores aggregated for each head, for both SBM and vanilla attention. Figure~\ref{fig:sbmatt} shows the two plots from the AST of the Python function in Listing~\ref{listing:code2}. With the vanilla attention, most of the nodes have high attention scores on a few specific nodes such as \emph{s} and \emph{upper}, generating vertical and horizontal patterns. SBM attention scores lack such pattern: the attention scores differ a lot between nodes. This result shows that SBM attention is capable of finding more specific attention scores. For example, while the identifier node of \emph{upper} attends a lot on \emph{s} and \emph{return\_statement}, the \emph{return} node attends on \emph{parameters} and the argument of the return statement, \emph{upper}. Additionally, the sparsity can improve the interpretability of the model, as there are fewer and more unique attention relationship between nodes.



\begin{figure*}[ht]
\subfigure[Python Forward Time \label{fig:forwardtime}]{\includegraphics[angle=0, width=4.3cm]{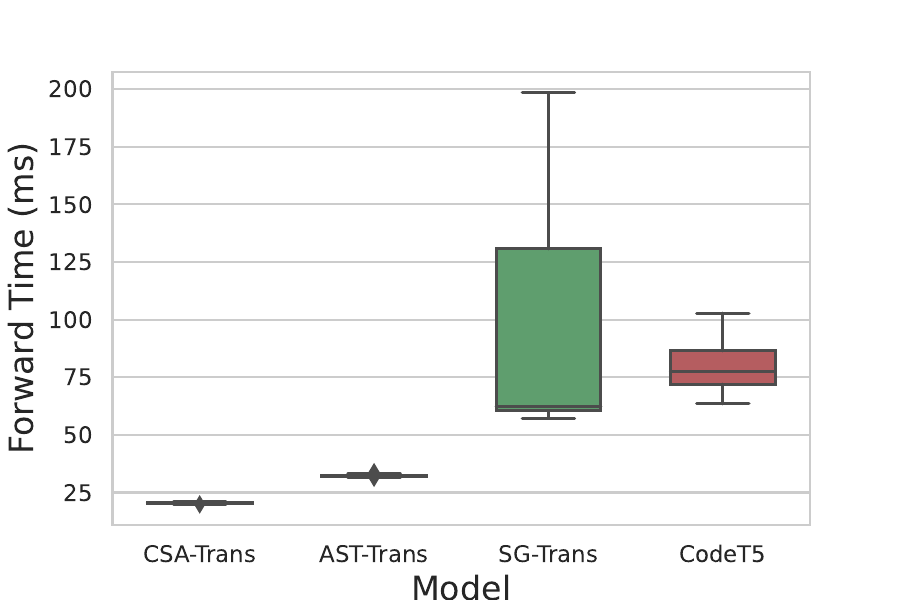}}
\subfigure[Python Forward Memory \label{fig:forwardmemory}]{\includegraphics[angle=0, width=4.3cm]{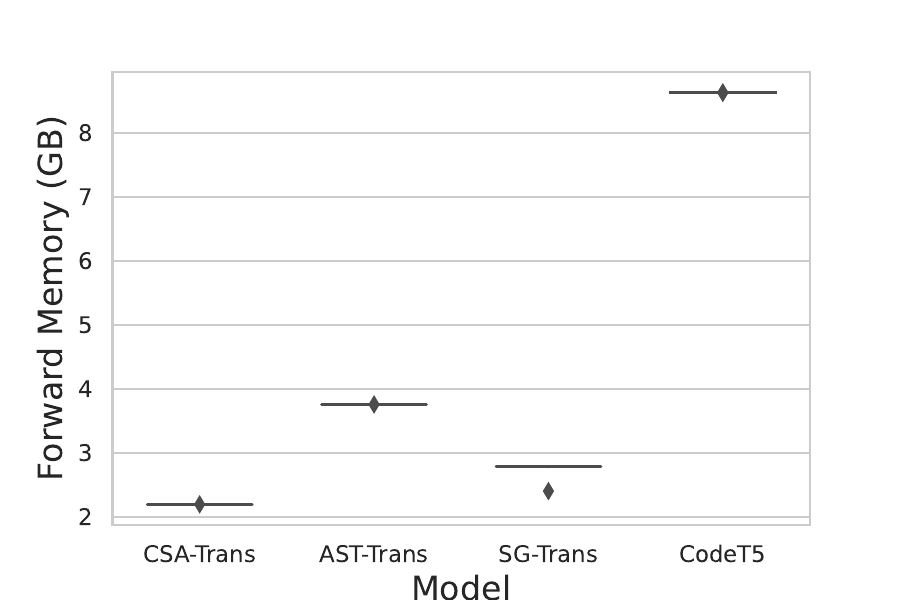}}
\subfigure[Python Backward Time \label{fig:backwardtime}]{\includegraphics[angle=0, width=4.3cm]{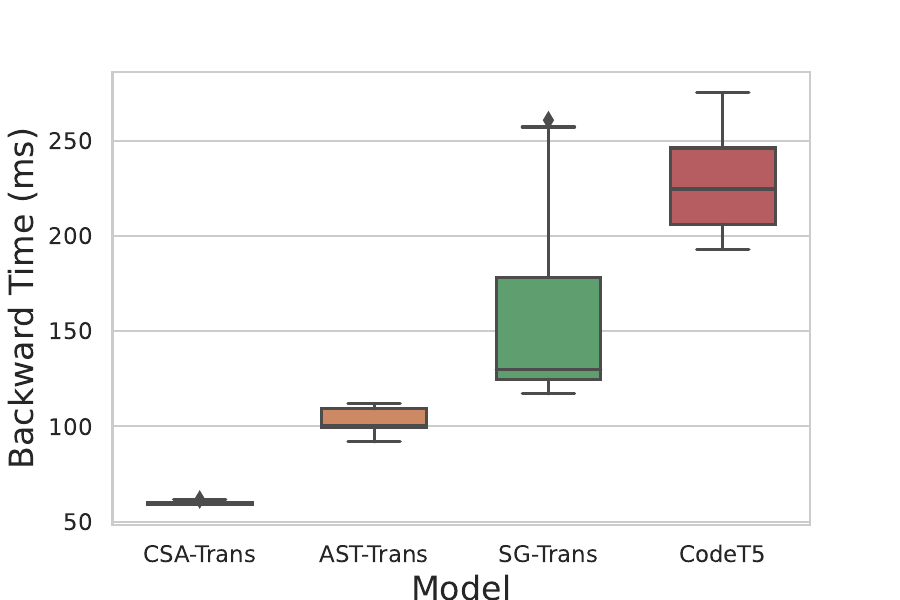}}
\subfigure[Python Backward Memory \label{fig:backwardmemory}]{\includegraphics[angle=0, width=4.3cm]{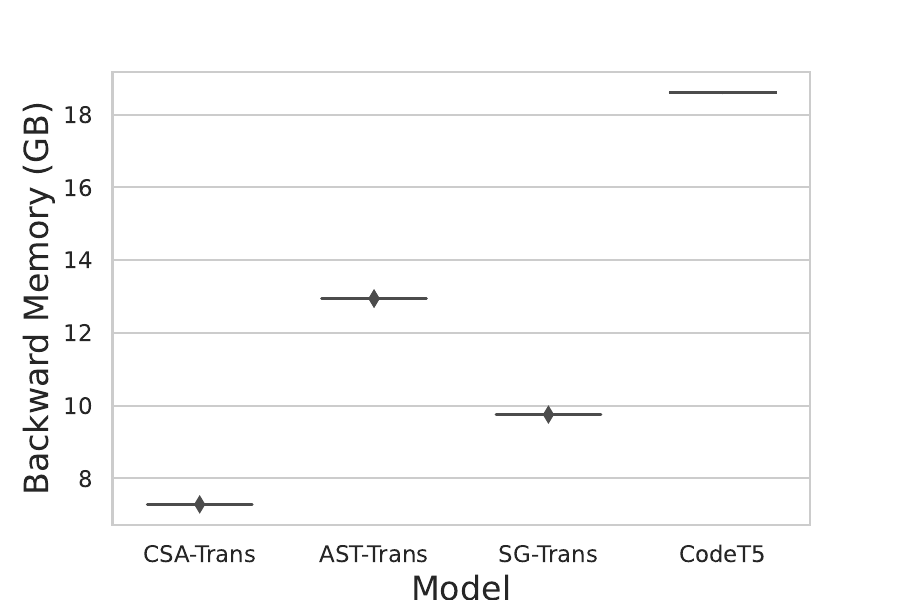}}

\subfigure[Java Forward Time \label{fig:javaforwardtime}]{\includegraphics[angle=0, width=4.3cm]{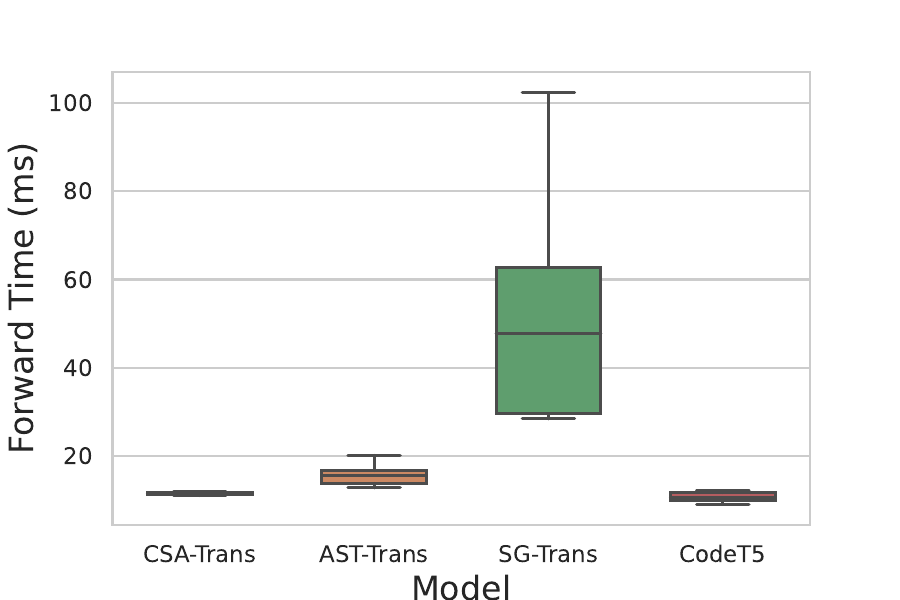}}
\subfigure[Java Forward Memory \label{fig:javaforwardmemory}]{\includegraphics[angle=0, width=4.3cm]{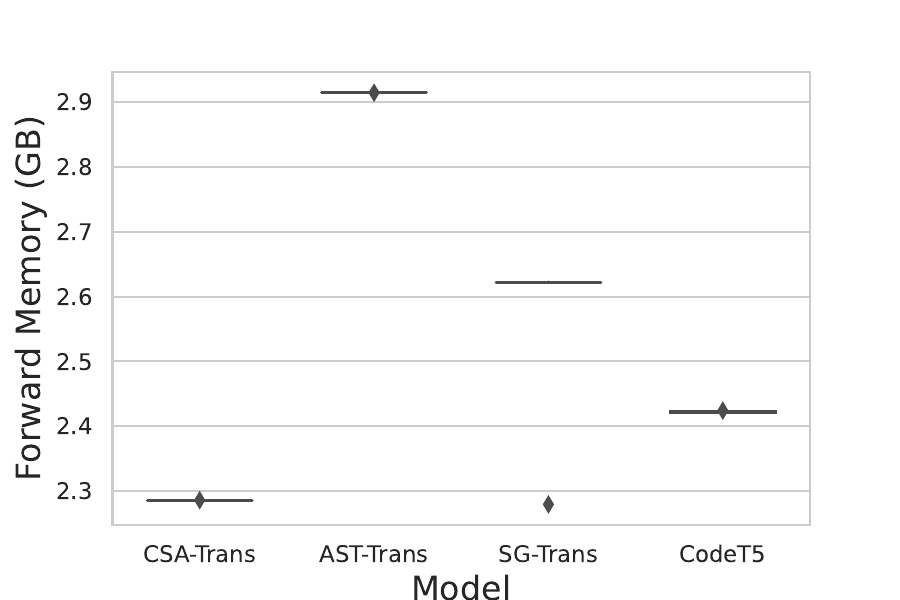}}
\subfigure[Java Backward Time \label{fig:javabackwardtime}]{\includegraphics[angle=0, width=4.3cm]{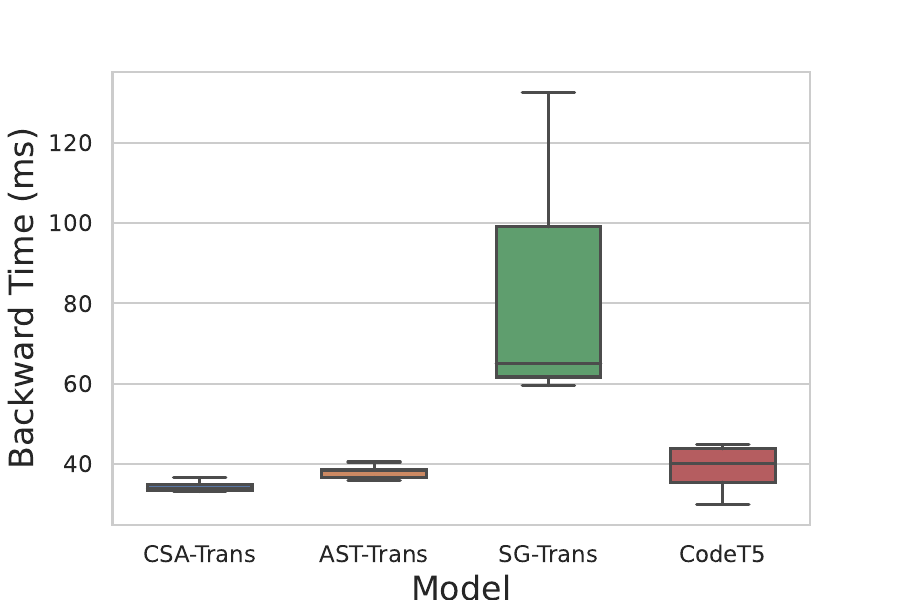}}
\subfigure[Java Backward Memory \label{fig:javabackwardmemory}]{\includegraphics[angle=0, width=4.3cm]{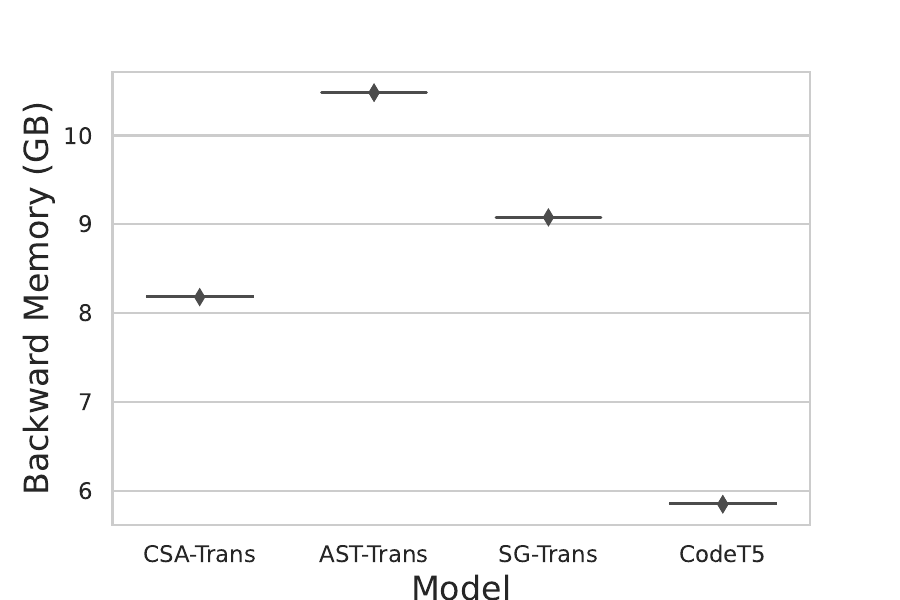}}

\caption{Time and Memory complexity}
\label{fig:complexity}
\end{figure*}

\subsection{RQ4: How efficient is CSA-Trans?}

To show the efficiency of CSA-Trans, we compare the time and memory consumption of CSA-Trans with three baselines, AST-Trans, SG-Trans, and CodeT5 in both Java and Python dataset. We measure the time and memory consumption with four criteria: forward/backward pass time and forward/backward peak memory usage. For each model, we record the time required by a forward pass (Forward Time). We then record the forward pass time followed by backpropagation (Backward Time). We then record the peak memory usage during the forward pass (Forward Memory) and backpropagation (Backward Memory). We repeat this experiment for 20 times and boxplot the result. 

Figure~\ref{fig:complexity} shows the backward time, and peak memory across the 20 runs. CSA-Trans is significantly more efficient than both AST-Trans and SG-Trans in both time and memory consumption. In terms of backward time CSA-Trans is 9.89\% faster than AST-Trans for Java and more surprisingly, 41.92\% faster for Python dataset. For memory consumption CSA-Trans consumes 9.83\% and 25.31\% less backward memory compared to SG-Trans for each Java and Python dataset. In terms of baselines, AST-Trans has lower forward, backward time compared to SG-Trans but requires much larger memory footprint. It is noteworthy that CSA-Trans outperforms both baselines for code summarization despite computationally being more efficient in both time and memory. Our simpler attention aggregation leads to leaner memory footprint compared to AST-Trans whose relative distance based attention aggregation results in heavy memory usage. Also, CSA-Trans does not require each encoder layer to use multiple harcoded masks, requiring reduced backward pass time.

In comparison to CodeT5, CodeT5 exhibits a 6.34\% improvement in forward time and a reduction of 28.49\% in backward memory usage. This discrepancy arises from the divergent attention mechanisms employed in CSA-Trans. CSA-Trans leverages techniques like disentangled and SBM attention, while CodeT5 employs simpler operations for attention computation. However, our findings in code summarization demonstrate that operations in CSA-Trans positively impact code summarization performance.

In contrast to prior results, CSA-Trans showcases better efficiency in both time and memory usage for the Python dataset. Also for Java dataset, CSA-Trans demonstrates a 12.53\% improvement in backward time and consumes 5.65\% less forward memory. This observed disparity can be attributed to the inherent properties of the aforementioned CodeT5 tokenizer, which requires more token outputs to fully build up the code summaries used in our evaluation.

\section{Related Work}
\label{sec:relatedwork}

\subsubsection{Code Summarization}
To help developer alleviate effot, different approaches based on templates~\cite{14McBurney}, IR~\cite{10Haiduc}, and machine learning~\cite{code2seq:19Alon,great:20Hellendoorn,tptrans:21Peng} have been proposed. Previous work representing code as sequential tokens use LSTM or Transformer architectures for embedding code~\cite{Iyer:CodeNN, Gao:sg-trans}. ASTs can be used in many different ways to extract information about source code. Code2Seq~\cite{code2seq:19Alon}, uses paths within ASTs, whereas other work use GNN and Transformer to embed ASTs themselves~\cite{Choi:Graph+Transformer, Liu:hybridmessagesummary}. Several recent work utilize both of the representations, either by aggregating code and AST embeddings, or by providing structural information obtained from AST to code tokens~\cite{guo:codescribe, great:20Hellendoorn}.

\subsubsection{Graph Neural Networks (GNN)} Graph is a data modality used to represent structural data used in different domains such as social networks~\cite{Qiu:socialnetwork}, recommendation systems~\cite{Wu:recommend}, and chemistry~\cite{Giler:chemistry}. Many of previous GNN work follow the message-passing scheme~\cite{Kipf:gcn, Velickovic:gat}, where each node embedding is updated by aggregating embeddings of its neighbors. Recently, GNNs capable of modeling long range interactions have been introduced~\cite{graphormer:21Ying, Maron:equivariant, Kim:higher-trans, Kim:tokengt}. Among them, \cite{graphormer:21Ying} showed that Transformer architecture can be applied to the graph domain. Recent code summarization models also adopted the Transformer architecture to embed ASTs~\cite{tptrans:21Peng, Tang:ast-trans}. CSA-Trans is aligned with this line of research, yet achieves state-of-the-art performance by introducing AST specific PE and the use of SBM attention.

\section{Limitations}
\label{sec:limitations}
Unlike hybrid approaches that both utilizes the AST and source code tokens, CSA-Trans only considers AST nodes for code summarization. While the rationale for using AST nodes comes from the fact that source codes are much more structured than natural languages, improvements using sequential PE~\cite{Brown:gpt3} suggests that the model may generate better summarization by incorporating both. Yet our work focuses on generating better summary using AST nodes and leave this as future work. 

One important setting when using AST is \emph{how} to generate source code AST. While we match settings with SG-Trans, we find that AST generation process varies across different baselines. For example, \cite{Choi:Graph+Transformer}, \cite{code2seq:19Alon} uses its own parser for Java codes while \cite{Hu:badcomment} uses Eclipse’s JDT compiler. While using different tools may generate slightly different ASTs, we were not able to reproduce all baselines using the same AST due to AST specific implementations. Yet, we use \emph{tree-sitter} for AST parsing which can be easily adopted, and use the same AST for replication of AST-Trans.

\section{Conclusion}
\label{sec:conclusion}

Self-attention is one of the most import components of the Transformer architecture. While various approaches have tried to embed ASTs using Transformer, they are limited in the way they capture structural information or the way they restrict self-attention calculation. To alleviate limitations, we present Code Structure Aware Transformer (CSA-Trans). CSA-Trans first learns the Positional Encoding for the given AST. Subsequently, using the learned PE, CSA-Trans applies SBM attention to allow global receptive field. An empirical comparison against 15 baselines shows that CSA-Trans can achieve better code summarization performance while being more efficient in terms of time and memory. Ablation studies of the learnt PE and SBM attention show that each component of CSA-Trans independently contribute to its performance. Quantitative and qualitative analysis of dynamically generated attention masks and scores reveals that dynamically generated attention scores are much more focused and sparser, suggesting better interpretability.

\bibliographystyle{IEEEtran}
\bibliography{custom}

\end{document}